\documentstyle[epsfig]{mn2e}
\def\msun{\rm {M}_{\odot}}

\def\simlt{\mathrel{\rlap{\lower 3pt\hbox{$\sim$}}\raise 2.0pt\hbox{$<$}}}
\def\simgt{\mathrel{\rlap{\lower 3pt\hbox{$\sim$}} \raise 2.0pt\hbox{$>$}}}
\def\lsim{\mathrel{\rlap{\lower 3pt\hbox{$\sim$}}\raise 2.0pt\hbox{$<$}}}
\def\gsim{\mathrel{\rlap{\lower 3pt\hbox{$\sim$}} \raise 2.0pt\hbox{$>$}}}

\begin{document}
\title[On the offset of SGRBs]{On the offset of Short Gamma-ray Bursts}
\author[Salvaterra et al.]{R.~Salvaterra$^1$, B.~Devecchi$^1$,
  M. Colpi$^2$, P. D'Avanzo$^3$\\$^1$ Dipartimento di Fisica \&
  Matematica, Universit\`a dell'Insubria, Via Valleggio 11, 22100
  Como, Italy\\ $^2$Dipartimento di Fisica G.~Occhialini, Universit\`a
  degli Studi di Milano Bicocca, Piazza della Scienza 3, 20126 Milano,
  Italy\\ $^3$INAF, Osservatorio Astronomico di Brera, via E. Bianchi
  46, 23807 Merate (LC), Italy}

\maketitle \vspace {7cm}

\begin{abstract}
Short Gamma-Ray Bursts (SGRBs) are expected to form from the
coalescence of compact binaries, either of primordial origin or from 
dynamical interactions in globular clusters. 
In this paper, we investigate the possibility that the
offset and afterglow brightness of a SGRB can help revealing the
origin of its progenitor binary. We find that a SGRB is likely to
result from the primordial channel if it is observed within 10 kpc
from the center of a massive galaxy and shows a detectable
afterglow. The same conclusion holds if it is 100 kpc away from a
small, isolated galaxy and shows a weak afterglow. On the other hand,
a dynamical origin is suggested for those SGRBs with observable 
afterglow either at a large separation from a massive,
isolated galaxy or with an offset of 10-100 kpc from a small,
isolated galaxy. We discuss the possibility that SGRBs from the dynamical
channel are hosted in intra-cluster globular clusters and find that 
GRB 061201 may fall within this scenario.
\end{abstract}


\section{Introduction}

The afterglows of several Short Gamma Ray Bursts (SGRBs) have recently
been localized on the sky with Swift (Gehrels et al. 2004), allowing
for a determination of their redshift and host galaxy (see
e.g. Berger 2009; Fong, Berger \& Fox 2010).  SGRBs are been observed
in all kinds of galaxies (from star-bursts to ellipticals and also
associated with galaxy clusters) with a wide range of offsets.

SGRBs are currently believed to result from the coalescence of a compact 
binary, either a double neutron star or a neutron star (NS)
and a black hole (BH) binary (Nakar 2007).  
Compact binaries are known to form in different environments along
two main channels: (i) from the evolution of massive stars in
primordial binaries rising in the galactic field (Narayan, Paczynski
\& Piran 1992), and (ii) through three (or a few)-body dynamical
interactions among stars and compact remnants in globular clusters
(GCs) (Grindlay, Portegies Zwart \& McMillan 2006).
Salvaterra et al. (2008) have shown that both formation channels are needed
in order to reproduce the {\it Swift} redshift distribution of SGRBs, 
with the dynamical channel in GCs contributing mainly at $z\simlt 0.3$.

In this paper, we investigate the nature of the observed offsets in
relation to the galaxy type and burst environment in order to
discriminate between the two channels and highlight the origin of
SGRBs.  A wide range of offsets is expected in both channels.  In the
primordial scenario, a large separation from the host galaxy can originate from
the natal kick rising at the time of formation of the compact object
(Belczynski et al. 2006). This mechanism is not present in case of
dynamical origin: few body interactions that are at the origin of
dynamical double neutron stars do not release binaries with large
recoil velocities (Devecchi et al. in preparation). So, the
offset in this latter case has to be ascribed to the underlying GC
spatial distribution. For isolated galaxies this reflects the GC
distribution that is known to decline more gently compared to stars
(see Brodie \& Strader 2006 and references therein).  In galaxy
clusters, GCs may have a wider spread, since there are hints (both
theoretical and observational) that a population of intra-cluster GCs
(ICGCs) can exist. SGRBs inside ICGCs can explain large potential
offsets in galaxy groups and clusters, besides natal kicks.

We outline here three different cases for the origin of the offsets:
(i) primordial SGRB kicked from a galaxy, isolated (Belczynski et
al. 2006) or in a cluster (Niino \& Totani 2008), by a natal kick;
(ii) dynamical SGRB in a GC bound to an isolated or cluster galaxy,
and (iii) dynamical SGRB in a ICGC.

\section{Quantifying the offset}

\subsection{Primordial SGRBs}

The theoretical spatial distribution of primordial SGRBs has been
computed for isolated galaxies of different types and 
sizes\footnote{Belczynski et al. (2006) considered three different galaxy
types, i.e. starburst, spiral and elliptical. They investigated both small
and large hosts with viral masses of $\sim 10^9\,\msun$ and $\sim 10^{12}\,\msun$, respectively.} using
population synthesis models by Belczynski et al. (2006).  The offset
results from the combination of the natal kick velocity with the time
that elapses from the formation of the compact binary and its
gravitational wave driven coalescence time. We select two windows for
the offset: the first between 0--10 kpc and the second between
10--100 kpc.  We note that with our definition of the offset we
include also SGRBs located well inside the host galaxy. From figure~3 of
Belczynski et al. (2006), we infer that the bulk of SGRB events
happens within the 10 kpc scale for star-bursts and spirals, whereas
for ellipticals this is true only for large hosts. For small ellipticals
only $\sim 15$\% shows this kind of offsets. The relative fraction of
SGRBs with small offset increases from early to late host galaxies, and from
small to large hosts. Instead, we find that the 10--100 kpc window is
always poorly populated by primordial SGRBs regardless the type and
size of the host galaxy, being the fraction of merging double neutron stars
(DNS) around $\sim 10-20$\%.  Finally, only for small ellipticals, we have a
good probability ($\sim 75$\%) to find SGRBs with very large offset
(i.e. $\simgt 100$ kpc) , whereas in other host types this fraction is
only $<5-20$\%.  This finding may not hold true if the host is member of
a galaxy cluster.  As shown by Niino \& Totani (2008), the fraction of
SGRB events occurring at very large offset may be as large as $\sim
80$\% if the potential well of each member galaxy is determined by
stars instead of dark matter due to dilution in the clustering
process. If, instead, the dark matter sub-halos are associated to member
galaxies as for field galaxies, the escape fraction
is only 20\%.

\begin{figure}
\begin{center}
\includegraphics[width=8cm]{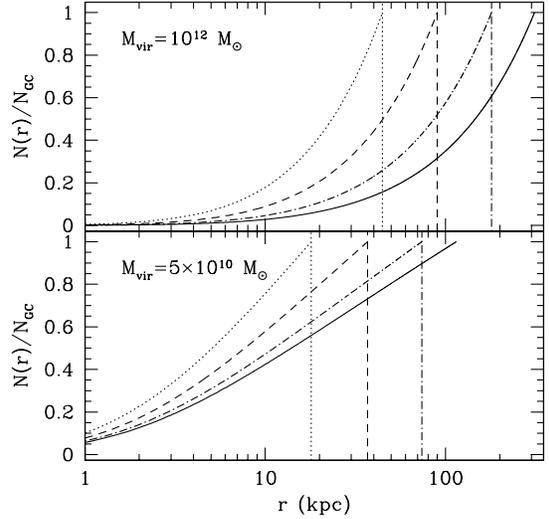}
\caption{Cumulative distribution of the GCs as a function of the
  projected distance $r$ normalized to the number of GCs in the galaxy.
 Top (bottom) panel refers to a  virial mass of $10^{12}\msun$ 
 ($5\times 10^{10}\msun$). Solid line refers to isolated
  galaxies. Other lines refer to host galaxies in Virgo-like cluster at 500 kpc
(dotted line), 1 Mpc (dashed line), and 2 Mpc (dot-dashed line) from
the cluster center. Vertical lines refer to the corresponding tidal
  radii.}
\end{center}
\end{figure} 

\subsection{Dynamical SGRBs: GCs bound to galaxy}

We consider a model for the GCs spatial distribution based on current
observations of extra-galactic GC spatial profiles. The number surface
density of the GC systems has been fitted in the literature either via
a power-law ($\Sigma\propto r^{-\alpha}$) or as a modified Hubble law
($\Sigma\propto (r^2+r_{\rm c}^2)^{-1}$). Trends between the
V-magnitude ($M_{\rm V}$) of the host galaxy and both $\alpha$ and
$r_c$ have been found, suggesting for a link between the evolution of
the GC systems and its host galaxy (Forbes et al. 1996; Ashman \& Zepf
1998): $ \alpha=0.28M_{\rm V}+7.5 $ and $r_{\rm c}=-0.62M_{\rm V}-11$
kpc.  In order to take into account both for the presence of a core
and the change in the slope of the outer profile, we here consider a
"mixed" model.  The number density of the GCs has been modeled as:

\begin{center}
\begin{equation}
  \Sigma(r)=\frac{\Sigma_{\rm 0}r^{\alpha-2}_{\rm c}}{\left(r+r_{\rm c}\right)^{\alpha}}.
\end{equation}
\end{center}

\noindent
The corresponding cumulative distribution is:

\begin{center}
\begin{equation}
  C(r)=\frac{2\pi\Sigma_{\rm 0}}{2-\alpha}\left[\left(\frac{r}{r_{\rm c}}+1\right)^{(1-\alpha)}\left(\frac{r}{r_{\rm c}}-\frac{1}{1-\alpha}\right)+\frac{1}{1-\alpha}\right]
\end{equation}
\end{center}

For each galaxy we relate $M_{\rm V}$ to the stellar mass by $M_{\rm V}=-2.5\log
M_\star +4.83$ and assume a ratio $M_\star/M_{\rm vir}=0.1$ between the
stellar and the dark matter mass. For an isolated galaxy, the
truncation radius for the GC distribution is taken to be the virial
radius. For a cluster galaxy the maximum radius at which the GCs can
still be bound to their host is the tidal radius $r_{\rm t}$ computed
assuming that each galaxy as well as the underlying cluster follows
the singular isothermal profile. For a galaxy at distance $R$ from the
cluster center, the tidal radius $r_{\rm t}=R \sigma_{\rm g}/\sigma_{\rm c}$, where
$\sigma_{\rm g}$ and $\sigma_{\rm c}$ correspond to the galaxy and cluster
velocity dispersion, respectively. We stress here that our models for
the GC spatial profile are based on the extrapolation of the observed
one up to the truncation radius.

The GC spatial distributions in large isolated host galaxies (i.e. $M_{\rm vir}\sim
10^{12}\;\msun$) are flatter that in smaller galaxies. GCs can be as
far as several hundred kpc, leading to offset of comparable extend.
In Fig.~1 we plot the
cumulative GC spatial distributions for two different galaxy
masses. Solid lines correspond to isolated systems. Dotted, dashed and
dot-dashed lines refer to galaxies located at 500 kpc, 1 Mpc and 2 Mpc
from the center of a Virgo--like cluster, respectively\footnote{We
  have assumed a mass for the Virgo cluster of $1.4\times 10^{15}\,
  \msun$ according to Fourqu{\'e} et al. (2001). This for an
  isothermal sphere corresponds to a $\sigma_{\rm c}\sim 1300\,$ km
  s$^{-1}$.}.  For galaxies harbored in clusters, tidal truncation
produces smaller offsets as shown in Fig.~1.  For the same offset
windows of Section 2.1, we find that the bulk of the GC population
residing in a large isolated host galaxy is expected to be outside the 100 kpc
scale\footnote{The existence of a possible break in the GC spatial
  distribution may reduce their fraction at very large distance.}.
The 10--100 kpc interval is also well populated ($\sim 30$\%),
contrary to the central ten kpc ($\sim 5$\%). In a small isolated host galaxy,
GCs are evenly distributed in the two windows.

For host galaxies in clusters the tidal truncation cuts the GC distribution so
that the 10--100 kpc is an ill defined window. In small galaxies, we
find the bulk of GCs inside 10 kpc. For the massive ones, less than
10-20\% of GC is in this offset window.  Bound GCs in large (small)
galaxies can extend out to 50-100 (18-75) kpc.

\subsection{Dynamical SGRBs: Intra-cluster GCs}

\begin{figure}
\begin{center}
\includegraphics[width=8cm]{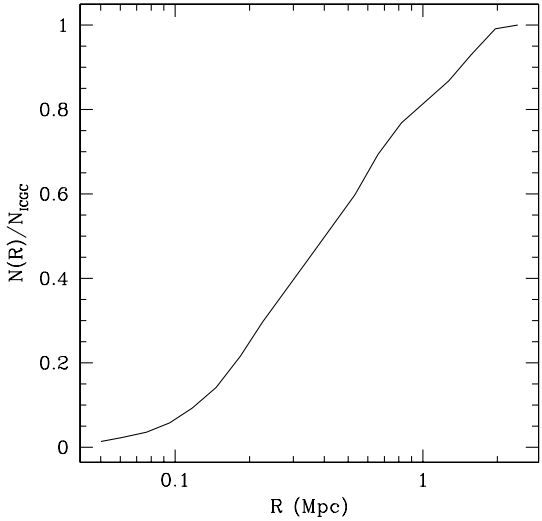}
\caption{Cumulative distribution of the ICGCs as a function of the
 projected distance $R$ from the center of the galaxy cluster, computed
using the projected radial density profile of figure~2 by Yahagi \& Bekki
(2005).}
\end{center}
\end{figure} 

Different observations (Bassino et al. 2002, 2003; Jord{\'a}n et
sl. 2003) indicate the existence of a population of ICGCs. Theoretical
investigations on the spatial distribution of GCs in galaxy clusters
predict an ICGC fraction of $\sim 30$\% regardless of the cluster
total mass (Bekki \& Yahagi 2006). ICGCs are spread over the cluster
volume and can be found far from the cluster center. Using the
projected radial density profile of figure~2 by Yahagi \& Bekki
(2005), we obtain the cumulative distribution of ICGCs as function of
the projected distance that is shown in Fig.~2. The plot illustrates
that ICGCs can be found up to very large distances from the cluster
center with $\sim 20$\% at $R>1$ Mpc.

\begin{figure*}
\includegraphics[width=12cm]{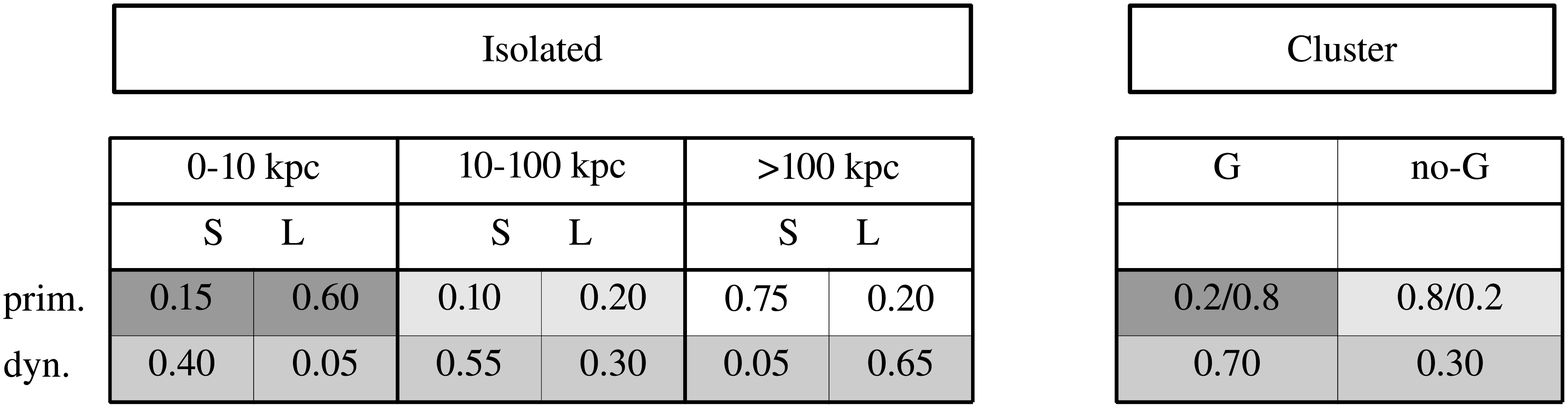}
\label{fig:riassunto}
\caption{Percentages of SGRB events with offset in the three domains
  for the two formation channels and the two galaxy models.  The gray
  scale indicates the afterglow visibility: darker colors refer to
  brighter afterglows, white is associated to a non detectable
  afterglow. The percentages are computed for SGRBs resulting from
  primordial binaries and from the dynamical channel. Left panel is
  for isolated galaxies where S(L) refers to the small (large) galaxy
  model.  Right panel is for SGRBs occurring in cluster of galaxies
  where G(no-G) refers to events bound (un-bound) to a cluster galaxy
  member. In the case of primordial galaxies there exist a large
  uncertainty ranging between 20-80\% for ejected SGRBs depending on
  dominance of dark matter in shaping the potential well of the member
  galaxies (see Niino \& Totani 2008).  Note that the table does not
  provide the relative contribution from the two channels but only the
  distribution in the three offset intervals.}
\end{figure*}

\section{Afterglow Detectability}

The intensity of the afterglow of GRBs is expected to be related to
the local environment around the burst (Sari, Piran \& Natarajan
1998).  Given the wide range of SGRB progenitors, of their offsets and
therefore of the diverse habitat of the explosion, we try here to
discuss possible constraints on the nature of the SGRB formation
channels from afterglow observations.  For the range of parameters and
observation times we are interested in, the afterglow can be modeled
as (slow cooling regime; Sari et al. 1998; Perna \& Belczynski 2002):

\begin{equation}
F_{\nu}\sim 1.1 n^{1/2} \xi_{\rm B}^{1/2} E_{50} d_{28}^{-2} (1+z)
(\nu/\nu_{\rm m})^{-2/3} \mbox{mJy}
\end{equation}

\noindent
where $\nu_{\rm m}=5.7\times 10^{13} \xi_{\rm e}^2 \xi_{\rm B}^{1/2}
E_{50}^{1/2} t_{\rm d}^{-3/2} (1+z)^{1/2} \mbox{Hz}$, $E_{50}$ is the
kinetic energy in units of $10^{50}$ erg s$^{-1}$, $d_{28}$ is the
luminosity distance in units of $10^{28}$ cm, and $t_{\rm d}$ is the
time from the explosion in days.  We assume here an adiabatic shock
and isotropic emission.  The $\gamma$-ray isotropic equivalent energy
output, $E_{\rm iso}$ is a reasonable estimator of $E$ and usually is
taken to be $10^{49-51}$ erg s$^{-1}$ (Nakar 2007).  $\xi_{\rm B}$ is
the fraction of the magnetic field energy density of the equipartition
value and $\xi_{\rm e}$ is the fraction of the internal density that
is carried by the electrons. $p$ is the power index of the electron
distribution. Typical values are $\xi_{\rm B}\sim 10^{-2.4}$,
$\xi_{\rm e}\sim 10^{-1.2}$ (derived for long GRBs; Panaitescu \&
Kumar 2001).

A SGRB originated from the primordial channel exploding at a very large offset 
(i.e. $\simgt 100$ kpc)  from an isolated host galaxy is embedded in the intergalactic
medium. Given
the very low gas density ($n\sim 10^{-7}$ cm$^{-3}$), the afterglow, if
present at all, should be probably too faint to be detected both in the
X-rays and in the optical. 

SGRBs from the primordial channel that blow well inside the host,
produce relative bright afterglows given the high density of the
interstellar medium.  Assuming a mean redshift for the primordial
population of $z\sim 0.5$ as suggested by Salvaterra et al. (2008), a
density $n\sim 1$ cm$^{-3}$ and $E_{50}=10$, the X-ray afterglow flux
is $\sim 2.7\times 10^{-10}$ erg s$^{-1}$ cm$^{-2}$ and detectable
with {\it Swift}/XRT. Optical observation by {\it Swift}/UVOT at 100
sec from the bursts can reveal the afterglow for the same parameters,
while 8-meter telescopes would detect it even one day after the
explosion with magnitude R$\sim 23.3$. Weaker explosion energies may
not provide enough signal to detect the optical afterglow.

For dynamically formed SGRBs, the burst originates inside the GC. The
gas density inside a GC has been measured only for 47 Tuc (Freire et
al. 2001, 2003) and is of the order of $\sim 0.1$ cm$^{-3}$. For a
SGRBs at an average $z=0.2$ (Salvaterra et al. 2008), one expect to
detect the X-ray afterglow for bursts with $E_{50}>0.5$. Also optical
detection with 8-m telescopes is possible for $E_{50}>1$. For
$E_{50}=10$, the X-ray flux is $5\times 10^{-10}$ erg s$^{-1}$ cm$^{-2}$
and magnitude R $\sim 22.6$ at 1 day from the burst. In this case, a
detection by the UV-optical telescope (UVOT) on aboard to {\it Swift}
may be possible.

The detectability of afterglows from primordial SGRBs ejected in the intra
cluster medium have been studied by Niino \& Totani (2008), that predicted
an observable X-ray afterglow as the burst is blowing in a medium of 
$\sim 10^{-3}$ cm$^{-3}$ for $z=0.2$ (Niino \& 
Totani 2008). The optical afterglow is in this case quite faint (magnitude R$\sim 26$
for a burst exploding at $z=0.2$ and observed at $t\sim 10^4$ s) requiring
8-m telescope follow-up observations. We note that SGRBs in ICGCs should
be brighter as they blow inside the denser medium of the GC, so that a 
possible discriminant could be the detection of a relative bright optical 
afterglow.

\section{Discussion}

In this paper, we showed how the combination of the observed spatial
offset, host type/mass and afterglow brightness can shed light on the
pathway of formation of SGRB binary progenitors, resulting either from
primordial stellar binaries, or from dynamical interactions in GCs. The
main results of our study are summarized in Fig.~3. Key results of
our analysis are:

\begin{itemize}
\item{If a SGRB is observed with a large separation from a massive,
  isolated galaxy we expect that it belongs to the dynamical channel
  and, residing in a globular cluster, it may show an observable X-ray
  and optical afterglow;}
\item{A SGRB exploding with a large offset from a small, isolated
  host galaxy is likely to ensue from the primordial channel. In this
  case the progenitor binary is ejected in the intergalactic medium
  following a large natal kick.  Nor or weak afterglow is expected
  given the very low density of the medium;}
\item{A SGRB observed within 10 kpc from the center of a massive galaxy
    (either isolated or in a cluster) should arise from the primordial channel
    with a detectable afterglow;}
\item{A SGRB associated to a galaxy cluster but not to any of its specific 
    members
    may result from both formation scenarios. A possible discriminant
    could be the detection of an optical afterglow that is expected to
    be brighter in the case of the denser intra-cluster globular cluster natal environment;}
\item{A SGRB with an offset of 10-100 kpc from a small, isolated
    galaxy is more probably originated from the dynamical channel and
    this may be confirmed from the identification of the afterglow.}
\end{itemize} 


We now apply our results to a few observed SGRBs. Troja et al. (2008)
suggested that SGRBs showing an extended-duration soft emission
component in their prompt emission preferentially have small projected
physical offsets.  Recent \textit{HST} observations of a sample of
SGRB host galaxies show that the distribution of offsets has a median
of $\sim 5$ kpc, about 5 times larger than for long GRBs, with no
evidence of differences between SGRBs with and without extended
emission (Fong et al. 2010).  In the case of a host galaxy at low redshift,
deep optical/NIR imaging can allow to resolve the galaxy surface
brightness profile; the detection of an optical afterglow can then
clearly pinpoint the SGRB position with respect to the host galaxy. This is the
case of GRB\,071227 ($z = 0.38$) and GRB\,060505 ($z =
0.09$). GRB\,071227 was firmly classified as a SGRB (Sato et al. 2007,
Golenetskii et al. 2007; Onda et al.  2008) and occurred on the plane
of a large ($r \sim 15$ kpc) spiral galaxy at a relatively large
offset ($\sim 15$ kpc; D'Avanzo et al 2009; Fong et al 2010). This
could equally favor both the primordial or dynamical formation channel
for its progenitor (Fig.~3). We note however that its location on the
galactic plane and within the light of the host likely favors a
primordial origin. The case of GRB\,060505 is even more interesting,
given that the classification of this GRB is still debated. The
duration ($T_{90} \sim 4$ s) and the spectral lag point towards a long
GRB classification (McBreen et al. 2008; Xu et al 2009), the
non-detection of an associated supernova down to deep limits favors for a
SGRB (Ofek et al. 2007), while detailed study of its host galaxy led to
different interpretations of its progenitor (Levesque \& Kewley 2007;
Thoene et al. 2008).  Nevertheless, we note that GRB\,060505 is an
outlier of the $E_{\rm p,i}-E_{\rm iso}$ correlation like all SGRBs of known
redshift and peak energy $E_{\rm p,i}$ (Amati et al.  2007). In the scenario
of a double compact object merger progenitor for GRB\,060505, the
position of the afterglow at an offset of 6.5 kpc from a large spiral
galaxy (Levesque \& Kewley 2007; Thoene et al. 2008), makes a
primordial origin highly probable (Fig.~3). A primordial origin is
also a valuable hypothesis for the progenitor of the farthest
short-duration ($T_{90} \sim 1.3$ s) GRB, occurred at $z \sim 2.6$
(GRB\,090426; Levesque et al. 2010; Antonelli et al. 2009). The
position of the optical afterglow inside the host galaxy, the
intrinsic absorption measured in the X-ray spectrum, and the redshift
could be indicative of a ``primordial'' binary system that merged in a
relatively short time ($10^7-10^8$ yr). On the other hand, a
core-collapse origin for this burst cannot be excluded in light of its
consistency with the $E_{\rm p,i}-E_{\rm iso}$ correlation that holds for long
GRBs (Antonelli et al.  2009), making the classification of
GRB\,090426 not straightforward.

As discussed above, the measure of the redshift of a SGRB is a key
issue to accurately evaluate its offset and to relate it with the size
of its host galaxy. However, some conclusions can also be drawn for
those SGRBs with no measured redshift. The detection of an optical
afterglow with sub-arsec precision\footnote{We note that most of the
  largest offsets inferred for SGRBs are based on X-ray positions,
  while only an optical (sub-arcsec) localization could give a firm
  GRB-galaxy association.} coincident with the profile of a galaxy
(see, e.g. GRB\,070707 and GRB\,051227; Piranomonte et al. 2008,
D'Avanzo et al. 2009) strongly hints for an association between the
two objects. For these bursts, offsets are consistent with zero, and
the SGRB positions follow the host light. From a statistical point of
view the majority of these systems probably originate from the
primordial channel as a population of SGRBs exploding inside GCs
should result in larger offsets.


A few interesting counter examples do exist.  GRB~061201 and
GRB~070809 are two SGRBs for which the detection of an optical
afterglow has allowed to determine their position with high
accuracy. Deep, follow-up observations fail to find a host galaxy
coincident with the optical afterglow. The field of GRB~061201 has
been studied by Stratta et al. (2007) with VLT and Fong et al. (2010)
with HST.  No host is revealed down to R(AB)$=26.1$ (Stratta et
al. 2007).  Two galaxies are in the field of GRB~061201: a spiral
galaxy at $z=0.111$ (Stratta et al. 2007) at a projected distance of
$\sim 32.5$ kpc and a fainter object with undetermined redshift at
1.8$^{\prime\prime}$ (Fong et al. 2010).  Although the ejection
hypothesis from one of these host candidates can not be excluded, we
note that the burst should have likely exploded in a gas poor
environment far away from the host. This is in contrast with the
relative bright afterglow observed. GRB~061201 is also know to be 0.9
Mpc away from the galaxy cluster ACO S 995 at $z=0.0865$. While the
ejection hypothesis from the cluster center appears unlikely due to
the high kick velocity required, we note that a sizable fraction
($\sim 30$\%, see Fig.~2) of the ICGC population is still present at
$R>0.9$ Mpc. We suggest that GRB~061201 originates from the dynamical
channel inside a ICGC of ACO S 995. The denser environment inside the
host GC may be responsible for the brightness of its optical
afterglow.

A second interesting case is GRB~070809 that is likely a SGRB (see
Barthelmy et al. 2007). Similar to GRB~061201, this burst shows an
optical afterglow but no underlying host galaxy to g(AB)=26.3 (Perley
et al. 2008).  A possible host candidate has been identified in a
small spiral galaxy ($\sim 2\times 10^{10}\;\msun$) at $z=0.2187$
(Perley et al. 2008)\footnote{Note the existence of another, very
  faint host candidate with undetermined redshift $\sim
  2.3^{\prime\prime}$ away from the optical afterglow position (Perley
  et al. 2008).}. At this redshift, the projected distance is $\sim
20$ kpc. Accordingly to Fig.~3 the probability of detecting a SGRB
with this offset from a small galaxy is rather low ($\sim 10$\%) for
the primordial channel. A dynamical origin is preferred in this case.
As for GRB~061201, the optical detection of the afterglow supports
this interpretation. We also notice that the dynamical channel is
expected to contribute most to the SGRB population at the redshift of
the putative host of GRB~070809 (Salvaterra et al. 2008).

These two examples show how the results presented in this paper can
provide a powerful tool to discriminate the origin of SGRBs.

\end{document}